\documentclass[twocolumn,floatfix,preprintnumbers,nofootinbib,superscriptaddress]{revtex4}

\usepackage{ulem}
\usepackage{bm}
\usepackage{times}
\usepackage{amssymb,amsbsy,amsmath,amsfonts}
\usepackage{graphicx}
\usepackage{float}
\usepackage{color}
\usepackage{morefloats}
\usepackage{rotating}
\usepackage{srcltx}
\usepackage{slashed}
\usepackage{subfigure}
\usepackage{multirow}
\usepackage{verbatim}
\usepackage{hyperref}
\usepackage{tabularx}
\usepackage{adjustbox}
\usepackage{array}
\usepackage{tabularray}
\begin{document}
	
\title{Theoretical study of $N(1535)$ and $\Sigma^*(1/2^-)$ in the Cabibbo-favored process $\Lambda_c^+ \to  p  \bar{K}^0\eta$}

\author{Ying Li}
\affiliation{School of Physics, Zhengzhou University, Zhengzhou 450001, China}

\author{Si-Wei Liu} 
\affiliation{Institute of Modern Physics, Chinese Academy of Sciences, Lanzhou 730000, China} 
\affiliation{School of Nuclear Sciences and Technology, University of Chinese Academy of Sciences, Beijing 101408, China}

\author{En Wang}~\email{wangen@zzu.edu.cn}
\affiliation{School of Physics, Zhengzhou University, Zhengzhou 450001, China}
\affiliation{Guangxi Key Laboratory of Nuclear Physics and Nuclear Technology, Guangxi Normal University, Guilin 541004, China}

\author{De-Min Li}~\email{lidm@zzu.edu.cn}
\affiliation{School of Physics, Zhengzhou University, Zhengzhou 450001, China}

\author{Li-Sheng Geng} \email{lisheng.geng@buaa.edu.cn}
\affiliation{School of Physics, Beihang University, Beijing 102206, China}
\affiliation{Beijing Key Laboratory of Advanced Nuclear Materials and Physics, Beihang University, Beijing, 102206, China}
\affiliation{Peng Huanwu Collaborative Center for Research and Education, Beihang University, Beijing 100191, China}
\affiliation{Southern Center for Nuclear-Science Theory (SCNT), Institute of Modern Physics, Chinese Academy of Sciences, Huizhou 516000, China}

\author{Ju-Jun Xie}~\email{xiejujun@impcas.ac.cn}
\affiliation{Institute of Modern Physics, Chinese Academy of Sciences, Lanzhou 730000, China} 
\affiliation{School of Nuclear Sciences and Technology, University of Chinese Academy of Sciences, Beijing 101408, China}
\affiliation{Southern Center for Nuclear-Science Theory (SCNT), Institute of Modern Physics, Chinese Academy of Sciences, Huizhou 516000, China}

\begin{abstract}

Motivated by the recent experimental measurements, we have investigated the Cabibbo-favored process $\Lambda_c^+ \to p \bar{K}^0\eta$, where the $N(1535)$ resonance is dynamically generated from the $S$-wave pseudoscalar meson-octet baryon interactions within the chiral unitary approach. The contributions from the intermediate $N(1650)$ and the predicted low-lying baryon $\Sigma^*(1/2^-)$ are also considered. In addition, a Breit-Wigner amplitude for the $N(1535)$ resonance is checked. By comparing with the measured $\eta p$, $\bar{K}^0 \eta$, and $p \bar{K}^0$ invariant mass squared distributions, our results support the interpretation of $N(1535)$ as a dynamically generated state. Furthermore, we demonstrate that, with the contribution from $\Sigma^*(1/2^-)$ taken into account, the calculated invariant mass spectrum agrees with the Belle measurements. Future precise measurements of the $\Lambda_c^+\to p  \bar{K}^0\eta$ process can further elucidate the existence of the low-lying baryon $\Sigma^*(1/2^-)$.

\end{abstract}
 
    \date{\today}
	\maketitle
	
\section{Introduction} \label{sec:Introduction}
Understanding the nature of the low-lying excited baryons with quantum numbers $J^P=1/2^-$ is one of the most challenging topics in hadron physics~\cite{Klempt:2007cp,Crede:2013kia}. Within the classical constituent quark model, one puzzle of the low-lying excited baryons is the ``mass reverse problem'',  i.e., the mass of the $N(1535)$ with spin-parity quantum numbers $J^P=1/2^-$ should be lower than the one of the radial excitation $N(1440)$ with $J^P=1/2^+$. It is also difficult to understand the strong coupling of the $N(1535)$ to the channels with strangeness, such as the $\eta N$ and $K\Lambda$ channels~\cite{ParticleDataGroup:2022pth,Liu:2005pm,Geng:2008cv}. Besides the strong couplings of the $N(1535)$ to the $\eta N$ and $K \Lambda$ channels, there is also evidence for a large effective $g_{N(1535) N \eta'}$ coupling from the $\gamma p \to p \eta'$ reaction~\cite{CLAS:2005rxe} and $pp \to pp \eta'$ reaction~\cite{Cao:2008st}.

In Refs.~\cite{Helminen:2000jb,Zou:2007mk,Zhang:2004xt,Hannelius:2000gu}, it was proposed that the $N(1535)$ resonance could be the lowest $L=1$ orbital excited $uud$ state with a large admixing of the $[ud][us]\bar{s}$ pentaquark component, which leads to a heavier $N(1535)$ than the $N(1440)$, and also gives a natural explanation for the large couplings of $N(1535)$ to the strangeness channels $\eta N$ and $K\Lambda$. In addition, $N(1535)$ could be dynamically generated from the $S$-wave pseudoscalar meson-octet baryon interactions within the chiral unitary approach and is predicted to strongly couple to the channels $\eta N$, $K\Lambda$, and $K\Sigma$~\cite{Kaiser:1995eg,Kaiser:1996js,Inoue:2001ip,Bruns:2010sv, Nieves:2011gb, Gamermann:2011mq, Khemchandani:2013nma}. When the pseudoscalar meson-baryon mixing with the vector-baryon states was considered, the physical picture remains unchanged~\cite{Garzon:2014ida}. The analyses of the $\phi$ production in $\pi p$ and $pp$ collisions suggest that the $N(1535)$ strongly couples to the $K\Sigma$, $K\Lambda$ and $\phi N$ channels~\cite{Xie:2007qt,Doring:2008sv}, which is consistent with the results of the chiral unitary approach. 
The $N(1535)$ was also interpreted as a three-quark core dressed by meson-nucleon scattering contributions within the Hamiltonian Effective Field Theory~\cite{Abell:2023nex,Guo:2022hud,Liu:2015ktc}.
We have proposed to study the $N(1535)$ within the chiral unitary approach through the three-body decays of $\Lambda_b$~\cite{Lyu:2023aqn,Lu:2016roh,Wang:2015pcn}. Recently, it was suggested to test the molecular nature of $N(1535)$ by measuring its correlation functions~\cite{Molina:2023jov}, or the scattering length and effective range of the channels $K\Sigma$, $K\Lambda$, and $\eta p$~\cite{Li:2023pjx}.

The multi-body non-leptonic decays of charmed baryons have proven to be a rich source of light hadrons since those processes have large phase space and involve complicated final-state strong interactions
~\cite{Oset:2016lyh,Miyahara:2015cja, Hyodo:2011js,Wang:2022nac,Xie:2017erh,Wang:2020pem,Zeng:2020och,Feng:2020jvp}.
Theoretically, the Cabibbo-favored process $\Lambda_c^+\to p \bar{K}^0 \eta$ had been proposed to test the molecular nature of the $N(1535)$ in Ref.~\cite{Xie:2017erh}, where the $\eta p$ invariant mass distribution was calculated with in the chiral unitary approach and an effective Lagrangian model. It was shown that the behavior of the $N(1535)$  in the two approaches is different in this process, and the $\eta p$ invariant mass distribution could be used to distinguish those two pictures of the $N(1535)$ resonance~\cite{Xie:2017erh}. Later, Ref.~\cite{Pavao:2018wdf} has re-analyzed this process by considering the mixing of pseudoscalar-baryon and vector-baryon interactions within the chiral unitary approach and found that both $N(1535)$ and $N(1650)$ could contribute to the $\Lambda_c^+\to p \bar{K}^0 \eta$ reaction. 

Recently, the Belle Collaboration measured the branching fraction $\mathcal{B}(\Lambda_c^+\to pK^0_S \eta)=(4.35\pm 0.10\pm 0.20\pm 0.22)\times 10^{-3}$, and reported the invariant mass squared distributions~\cite{Belle:2022pwd}. In the $\eta p$ invariant mass squared distribution, one finds a narrow peak around $M^2_{\eta p}=2.3$~GeV$^2$ and a broad peak around $M^2_{\eta p}=2.75$~GeV$^2$, which should correspond to the $N(1535)$ and $N(1650)$ resonances, respectively. In addition, a threshold enhancement around  $M^2_{ p{K}_S^0}=2.1$~GeV$^2$ in the $p{K}_S^0$ invariant mass squared distribution could be associated with the predicted low-lying excited $\Sigma^*(1/2^-)$ baryon state below the $pK^0_S$ threshold. To date, the $\Sigma^*(1/2^-)$ state has not yet been well confirmed experimentally and theoretically, and searching for the $\Sigma^*(1/2^-)$ state is crucial for deepening our understanding of low-lying excited baryons~\cite{Roca:2013cca,Wu:2009tu,Wu:2009nw,Gao:2010hy,Xie:2017xwx,Xie:2018gbi,Liu:2017hdx,Wang:2015qta,Lyu:2023oqn,Kim:2021wov,Lu:2022hwm}. Thus, the Belle measurements of the process $\Lambda_c^+\to pK^0_S \eta$ could be useful to explore the nature of $N(1535)$ and to search for the predicted low-lying $\Sigma^*(1/2^-)$ state.

In this work, we investigate the process $\Lambda_c^+ \to p \bar{K}^0\eta$ by considering the contributions from $N(1535)$ in the chiral unitary approach and the Breit-Wigner parameterization (in this way, we take $N(1535)$ as a genuine baryon state). In addition, the intermediate resonance $N(1650)$ and the predicted low-lying baryon $ \Sigma^*(1/2^-)$ are also considered. We want to distinguish the molecular picture and genuine baryon state for the $N(1535)$ by comparing our calculations with the Belle measurements. At the same time, we demonstrate that, with the contribution from $\Sigma^*(1/2^-)$, the calculated invariant mass distributions agree well with the experimental measurements.

The structure of this paper is as follows. The theoretical framework is presented in Sec.~\ref{sec:Formalism}.  Results and discussions are presented in Sec.~\ref{sec:Results}. Finally, a summary is given in Sec.~\ref{sec:Conclusions}.
	
\section{Formalism} \label{sec:Formalism}

In Subsec.~\ref{susec:A}, we first present the theoretical formalism for the Cabibbo-favored process $\Lambda_c^+ \to p  \bar{K}^0\eta$, where the $N(1535)$ is dynamically generated from the $S$-wave pseudoscalar meson-octet baryon interactions. For comparison, we also adopt the Breit-Wigner form for the $N(1535)$ contribution by regarding the $N(1535)$ as a genuine baryon state. In Subsec.~\ref{susec:B}, we present the contributions from the $N(1650)$ and the predicted low-lying baryon $\Sigma^*(1/2^-)$. Finally, we present the formalism for the invariant mass squared distributions of the process $\Lambda_c^+ \to p \bar{K}^0\eta$ in Subsec.~\ref{susec:C}.

\subsection{$S$-wave meson-baryon interactions and $N(1535)$}\label{susec:A}

\begin{figure}[htbp]
		\centering
			\includegraphics[scale=0.6]{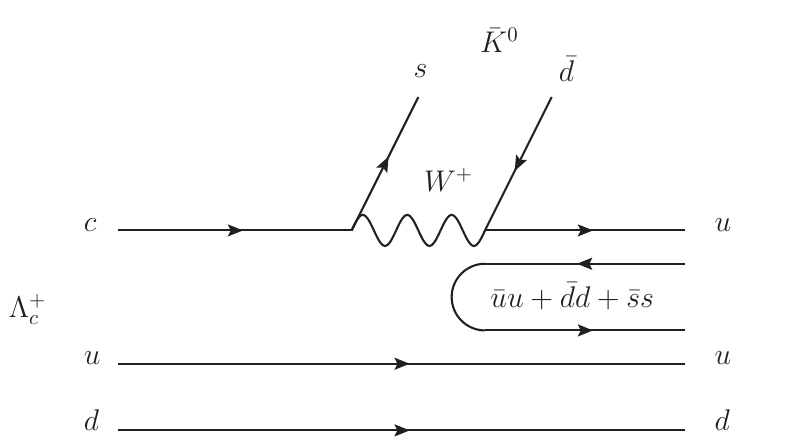}
		\caption{The quark-level diagram for the weak decay of $\Lambda_c^+$.}
		\label{fig:quarklevel}
	\end{figure}

As done in Refs.~\cite{Miyahara:2015cja,Wang:2022nac,Xie:2017erh}, for the Cabibbo-favored process $\Lambda_c^+\to \bar{K}^0 p \eta$, the dominant mechanism is the $W^+$ internal emission, as depicted in Fig.~\ref{fig:quarklevel}. First, the $c$ quark of the initial $\Lambda_c^+$ weakly decays into an $s$ quark and a $W^+$ boson; then the $W^+$ boson subsequently decays into a $u\bar{d}$ quark pair. Secondly, the $s$ quark and the $\bar{d}$ quark hadronize into  $\bar{K}^0$, while the $u$ quark from the decay of the $W^+$ boson and the $ud$ pair of the initial $\Lambda_c^+$, together with the antiquark-quark pair $\bar{q}q (=\bar{u}u+\bar{d}d+\bar{s}s)$ created from the vacuum with the quantum numbers $J^{PC}=0^{++}$, hadronize into hadron pairs. This could be expressed as,
     \begin{align}
\Lambda_c^+&\Rightarrow \frac{1}{\sqrt{2}}c\left(ud-du\right) \Rightarrow \frac{1}{\sqrt{2}}sW^+\left(ud-du\right)\nonumber\\
           &\Rightarrow \frac{1}{\sqrt{2}}s\bar{d}u\left(\bar{u}u+\bar{d}d+\bar{s}s\right)\left(ud-du\right)\nonumber\\
           &\Rightarrow \frac{1}{\sqrt{2}}\bar{K}^0u\left(\bar{u}u+\bar{d}d+\bar{s}s\right)\left(ud-du\right)\nonumber\\
           &\Rightarrow \frac{1}{\sqrt{2}}\bar{K}^0\sum M_{1i}q_i\left(ud-du\right),
           \label{eq:MBchannel}
\end{align}

with the $q\bar{q}$ matrix defined as,
\begin{eqnarray}
		M =\left(\begin{matrix}  u\bar{u}  & u\bar{d}  & u\bar{s}  \\
			d\bar{u}  &   d\bar{d}  &  d\bar{s} \\
			s\bar{u}  &  s\bar{d}   & s\bar{s}
		\end{matrix}
		\right),
	\end{eqnarray}
which can be expressed in terms of pseudoscalar mesons as~\cite{Lyu:2023ppb,Li:2023nsw},
\begin{eqnarray}
		M =\left(\begin{matrix}  \frac{{\eta}}{\sqrt{3}}+\frac{{\pi}^0}{\sqrt{2}} +\frac{{\eta}'}{\sqrt{6}}  & \pi^+  & K^{+}  \\
			\pi^-  &    \frac{{\eta}}{\sqrt{3}}- \frac{{\pi}^0}{\sqrt{2}}+\frac{{\eta}'}{\sqrt{6}}  &  K^{0} \\
			K^{-}  &  \bar{K}^{0}   &   \sqrt{\frac{2}{3}}{\eta}'-\frac{{\eta}}{\sqrt{3}}
		\end{matrix}
		\right),
	\end{eqnarray}
where we have taken the $\eta$-$\eta'$ mixing from Refs.~\cite{Bramon:1992kr}. 

Then, we can obtain the relevant components of the meson-baryon pair as,
 \begin{align}
           &\frac{1}{\sqrt{2}}\sum M_{1i}q_i\left(ud-du\right)\nonumber\\
           =&\left(\frac{1}{\sqrt{2}}\pi^0+\frac{1}{\sqrt{3}}\eta\right)\left[\frac{1}{\sqrt{2}}u\left(ud-du\right)\right]\nonumber\\
           &+\pi^+\left[\frac{1}{\sqrt{2}}d\left(ud-du\right)\right]+K^+\left[\frac{1}{\sqrt{2}}s\left(ud-du\right)\right]\nonumber\\
           =&\frac{\sqrt{2}}{2}\pi^0p+\frac{\sqrt{3}}{3}\eta p+\pi^+n-\frac{\sqrt{6}}{3}K^+\Lambda,
\end{align}
where the $\eta'p$ channel is ignored since its mass threshold $(1896~ \mathrm{MeV})$ is far from the mass region of $N(1535)$. Here, we take the flavor-wave functions for the octet baryons as~\cite{Pavao:2017cpt,Miyahara:2016yyh,Lyu:2023aqn}, 
	\begin{gather}
		p=\frac{u(ud-du)}{\sqrt{2}}, ~~~~~~	n=\frac{d(ud-du)}{\sqrt{2}} ,\nonumber\\
		\Lambda=\frac{u(ds-sd)+d(su-us)-2s(ud-du)}{2\sqrt{3}}.
	\end{gather} 

In addition, with the isospin triplet $(-\pi^+, \pi^0, \pi^-)$ and isospin doublet $(p, n)$~\cite{Close:1979bt}, we obtain, 
\begin{eqnarray}
\frac{\sqrt{2}}{2}|\pi^0p\rangle+|\pi^+n\rangle =-\frac{\sqrt{6}}{2}|\pi N\rangle^{I=1/2}.
\end{eqnarray}

With the above ingredients, the components of the pseudoscalar meson-octet baryon in Eq.~(\ref{eq:MBchannel}) could be rewritten as,

\begin{eqnarray}
\Lambda_c^+\Rightarrow \bar{K}^0\left(-\frac{\sqrt{6}}{2}\pi N+\frac{\sqrt{3}}{3}\eta N-\frac{\sqrt{6}}{3}K \Lambda  \right).
\end{eqnarray}

		\begin{figure}[htbp]
		\centering
		\subfigure[]{
			\includegraphics[scale=0.8]{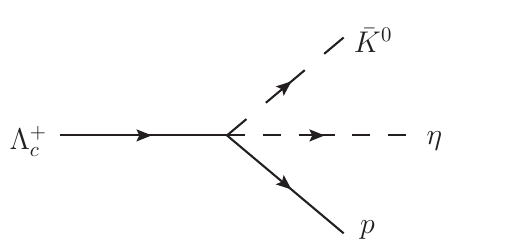}
   }
		\subfigure[]{
			\includegraphics[scale=0.8]{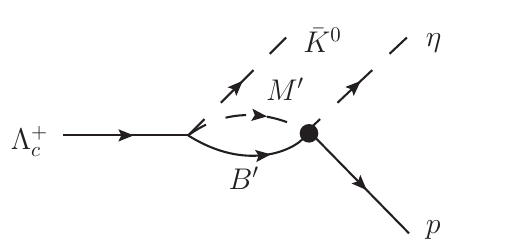}
   }
		\caption{
  Diagrams for the process $\Lambda_c^+ \to p \bar{K}^0\eta$. (a) Tree diagram, (b) final-state interactions of the pseudoscalar mesons and octet baryons.}
		\label{fig:FSI}
	\end{figure}
 
Thus, the process $\Lambda_c^+\to p\bar{K}^0 \eta$ could occur via the tree diagram of Fig.~\ref{fig:FSI}(a) and the $S$-wave pseudoscalar meson-octet baryon interactions of Fig.~\ref{fig:FSI}(b). The corresponding decay amplitude is given by,
\begin{eqnarray}
    \mathcal{T}^{N(1535)}&=& V_1 \left(h_{\eta N} + h_{\eta N}G_{\eta N}t_{\eta N \to \eta N} \right. \nonumber  \\
    && \!\!\!\! \!\!\!\! \left. + h_{\pi N}G_{\pi N}t_{\pi N \to \eta N}  + h_{K \Lambda}G_{K \Lambda}t_{K \Lambda \to \eta N}\right), \label{eq:amp_1535_BS}
\end{eqnarray}
with the coefficients
\begin{equation}
 h_{\eta N}=\frac{\sqrt{3}}{3},~
h_{\pi N}=-\frac{\sqrt{6}}{2},~
h_{K\Lambda}=-\frac{\sqrt{6}}{3},
\end{equation}
Here, the $V_1$ stands for the strength of the weak production vertex and could be treated as a constant~\cite{Miyahara:2015cja,Feng:2020jvp,Lu:2016ogy}.\footnote{
In obtaining this decay amplitude, we have assumed the factorization of the hard process (the weak decay and the hadronization) and the following rescattering of hadrons.  In this work, the coupling for the weak vertex of Fig.~2 is approximated as a constant, the only thing demanded is that this factor is smooth and practically constant as a function
of the invariant masses in the limited range where the predictions are made. This factorization works well in the previous studies~\cite{Geng:2018upx,Xie:2016evi}.
}. The $G_i$ is the meson-baryon loop function and $t_{i \to \eta N}$ is the transition amplitude between the $i$-channel and $\eta N$. These transition amplitudes are obtained by solving the Bethe-Salpeter equation,
\begin{equation}
    T=[1-VG]^{-1}V,
   \end{equation}
where $V$ is the interaction kernel matrix between coupled channels, which are $\pi N$, $\eta N$, $K\Lambda$, and $K\Sigma$. 

The transition potential $V_{ij}$ can be obtained from the lowest order chiral Lagrangian~\cite{Inoue:2001ip,Wang:2015pcn},
\begin{eqnarray}
	V_{ij}&=&-C_{ij}\frac{1}{4f_if_j}\left(2\sqrt{s}-M_i-M_j\right) \nonumber \\
  && \times \sqrt{\frac{E_i+M_i}{2M_i}}\sqrt{\frac{E_j+M_j}{2M_j}},
	\end{eqnarray} 
where the coefficients $C_{ij}$, reflecting the SU(3) flavor symmetry, are shown in Table~\ref{tab:cij}. In this work, we take the decay constants $f_{\pi}=93$~MeV, $f_K=1.22f_{\pi}$, and $f_{\eta}=1.3f_{\pi}$. $E_i(E_j)$ is the energy of the baryon in the $i$-th channel, 
 	\begin{align}
  E_{i(j)}=\frac{s+M_{i(j)}^2-m_{i(j)}^2}{2\sqrt{s}},
	\end{align} 
where $M_{i}$ and $m_{i}$ are the  masses of the baryon and meson in the $i$-th channel, respectively, and $s=M^2_{MB}$ is the invariant mass squared of the meson-baryon system. 

\begin{table}[htbp]
	\caption {$C_{ij}$ coefficients in the potential ($C_{ij}=C_{ji}$).
	} \label{tab:cij}
	\begin{tabular}{lcccc}
		\hline\hline  
		\qquad\quad&$\pi N$\qquad\quad&$\eta N$\qquad\quad&$K\Lambda$\qquad\quad&$K\Sigma$   \\  \hline 
		$\pi N$\qquad\qquad&2\qquad\qquad &0 \qquad\qquad &$3/2$\qquad\quad&$-1/2$\\ $\eta N$\qquad\qquad& \qquad\qquad &0\qquad\qquad&$-3/2$ \qquad\quad&$-3/2$    \\
		$K\Lambda$\qquad\qquad& \qquad\qquad &  \qquad\qquad &0\qquad\quad&0\\
		$K\Sigma$\qquad\qquad& \qquad\qquad &  \qquad\qquad & \qquad\quad&2\\
		\hline\hline
	\end{tabular}
\end{table}

The meson-baryon loop functions $G_i$  could be expressed as~\cite{Inoue:2001ip},
 \begin{eqnarray} \label{eq:Gfunction}
     G_i(s) =i\int \frac{d^4q}{(2\pi)^4}\frac{2M_i}{(P-q)^2-M_i^2+i\epsilon}\frac{1}{q^2-m_i^2+i\epsilon},
 \end{eqnarray}
where $P$ and $q$ are the four-momentum of the meson-baryon system and the meson, respectively, and $\sqrt{s}$ is the invariant mass of the meson-baryon system $(s=P^2)$. The integral is logarithmically divergent, and usually, one can use the cut-off method or the dimensional regularization to solve this singular integral~\cite{Inoue:2001ip,Guo:2005wp}.
In the numerator of the baryon propagator of Eq.~(\ref{eq:Gfunction}), a nonrelativistic approximation is utilized. Since the three-momenta of the baryon in the coupled-channels is much smaller than the baryon mass,  it is anticipated that this approximation would not affect the our results significantly.
Here, we calculate the integral using the three-momentum cut-off method, and the analytic expression of the loop function $G_i$ could be expressed as~\cite{Guo:2005wp},	
	\begin{align}
		G(s) =&\frac{2M}{16\pi^2s}\left\lbrace\sigma\left({\rm arctan}\frac{s+\Delta}{\sigma\lambda_1}+{\rm arctan}\frac{s-\Delta}{\sigma\lambda_2}\right) \right.\nonumber\\
  &-\left[(s+\Delta){\rm ln}\frac{(1+\lambda_1)q_{\rm max}}{m_1} \right. \nonumber \\
  &\left. \left. +(s-\Delta){\rm ln}\frac{(1+\lambda_2)q_{\rm max}}{m_2}\right]\right\rbrace,\label{eq:loopfunction}
	\end{align} 
where $\sigma$, $\Delta$, $\lambda_1$, and $\lambda_2$ are given as follows,
	\begin{gather}
	\sigma=\left[-\left(s-(M_i+m_i)^2\right)\left(s-(M_i-m_i)^2\right)\right]^{1/2} ,\\
   \Delta=M_i^2-m_i^2 ,\\
	\lambda_1 = \sqrt{1+\frac{M_i^2}{q^2_{\rm max}}}, ~
	\lambda_2 = \sqrt{1+\frac{m_i^2}{q^2_{\rm max}}}.
	\end{gather} 
Here $q_{\rm max}$ is a three-momentum cut-off parameter. In this work, we take $q_{\rm max}$ = 1150~MeV  to produce the pole position of the dynamically generated $N(1535)$ at the one quoted in the Review of Particle Physics (RPP)~\cite{ParticleDataGroup:2022pth}. 
The pole position for the dynamically generated $N(1535)$ is (1509.43 + 34.08i) MeV. In the complex plane, the loop function in the second Riemann sheet can be expressed from the one in the first Riemann sheet by:
        \begin{eqnarray}
        G^{II}(\sqrt{s})=G^I(\sqrt{s})+i\frac{|\vec{q}|}{4\pi\sqrt{s}},\qquad \mathrm{Im}(|\vec{q}|)>0,
        \end{eqnarray}
       where $\vec{q}$ is the momentum of the meson in the centre of mass frame. When searching for poles we use $G^I$ for $(\sqrt{s}) < m_1+m_2$, and $G^{II}$ for $(\sqrt{s}) > m_1+m_2$.

To distinguish between the molecular and the genuine baryon pictures, we also adopt the Breit-Wigner amplitude from Ref.~\cite{BESIII:2013xkm} for the contribution of the $N(1535)$ to explore its genuine baryon state picture,\footnote{
In fact, this amplitude should include the Fermion propagator of $\gamma_\mu p^\mu + M_{N^*}$. However, the three-momentum of the $N(1535)$ around its peak is much smaller than its energy. As a result, we have ignored these contributions from its small three-momentum terms, and taken the simple Breit-Wigner form for the contribution from $N(1535)$ resonance [also for the case of $N(1650)$].}
	\begin{equation}
	\tilde{\mathcal{T}}^{N(1535)} = \frac{\tilde{V}_1M_{N(1535)}\Gamma_{N(1535)}^0}{M^2_{\eta p}-M^2_{N(1535)}+iM_{N(1535)}\Gamma_{N(1535)}(s)},  \label{eq:amp_1535_BW}
	\end{equation}
with the phase-space-dependent width for $N(1535)$ which can be written as~\cite{Liu:2005pm,BESIII:2013xkm,Xie:2017erh} 
\begin{align}
       \Gamma_{N^*}(s) = \Gamma_{N^*}^0 \left( 0.5 \frac{\rho_{N\pi}(s)}{\rho_{N\pi}(M_{N^*}^2)} + 0.5 \frac{\rho_{N\eta}(s)}{\rho_{N\eta}(M_{N^*}^2)} \right),
  \label{eq:amp_1535_BW_gamma}
	\end{align} 
where $\rho_{N\pi}$ and $\rho_{N\eta}$ are the phase space factors for the final states.
The mass and width of $N(1535)$ are $M_{N(1535)}=1530$~MeV and $\Gamma_{N(1535)}^0=150$~MeV~\cite{ParticleDataGroup:2022pth}, respectively, and $\tilde{V}_1$ is the relative weight of the $N(1535)$ contribution.

\subsection{Contributions from $N(1650)$ and $\Sigma^*(1/2^-)$}\label{susec:B}

\begin{figure}[htbp]
		\centering
			\includegraphics[scale=0.8]{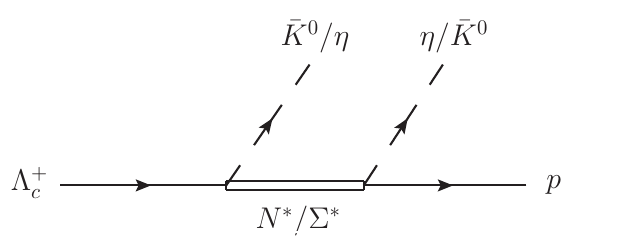}
		\caption{Tree-level diagram accounting for the contribution from the intermediate $N(1650) (N^*)$ and $\Sigma^*(1/2^-) (\Sigma^*$).}
		\label{fig:highN}
	\end{figure}

As discussed in the introduction, besides the $N(1535)$, one could find a bump structure around $M_{\eta p}^2=2.75$~GeV$^2$ in the $\eta p$ invariant mass distribution of the Belle measurements, which can be associated with the $N(1650)$. Furthermore, there is a threshold enhancement around $M_{p{K}_S^0}^2=2.1$~GeV$^2$ in the $p{K}_S^0$ invariant mass squared distribution, which may be associated with the predicted low-lying baryon $\Sigma^*(1/2^-)$. Hence, we also consider the contributions from the resonances $N(1650)$ and $\Sigma^*(1/2^-)$ in the processes 
$\Lambda^+_c\to \bar{K}^0 N(1650) \to p\bar{K}^0\eta$ and $\Lambda^+_c\to \eta \Sigma^*(1/2^-) \to p\bar{K}^0\eta$, as depicted in Fig.~\ref{fig:highN}. The Breit-Wigner decay amplitudes of Fig.~\ref{fig:highN} can be written as,
	\begin{align}
\mathcal{T}^{N(1650)}&=\frac{V_2M_{N^*}\Gamma_{N^*}}{M^2_{\eta p}-M^2_{N^*}+iM_{N^*}\Gamma_{N^*}}, \\
 \mathcal{T}^{\Sigma^*(1/2^-)}&=\frac{V_3M_{\Sigma^*}\Gamma_{\Sigma^*}}{M^2_{p\bar{K}^0}-M^2_{\Sigma^*}+iM_{\Sigma^*}\Gamma_{\Sigma^*}},
	\end{align} 
where the parameters $V_2$ and $V_3$ are the relative weights of the contributions from the intermediate resonances $N(1650)$  and $\Sigma^*(1/2)^-$, respectively, which will be determined by fitting the Belle measurements~\cite{Belle:2022pwd}. The mass and width of $N(1650)$ are $M_{N(1650)}=1665$~MeV and $\Gamma_{N(1650)}=125$~MeV~\cite{ParticleDataGroup:2022pth}, respectively. Furthermore, we have studied the process of $\Lambda_c^+\to \eta \Lambda \pi^+$ and found that the Belle measurements favor the existence of a $\Sigma^*(1/2^-)$ with a mass about 1380~MeV and width about 120~MeV in the intermediate process  $\Lambda_c^+\to \eta\Sigma^*(1/2^-) \to\eta \Lambda \pi^+$~\cite{Lyu:2024qgc,Belle:2020xku}. Thus, in this study we take $M_{\Sigma^*(1/2^-)}=1380$~MeV and $\Gamma_{\Sigma^*(1/2^-)}=120$~MeV, as used in Refs.~\cite{Wu:2009tu,Wu:2009nw,Liu:2017hdx,Lyu:2024qgc}.

It should be stressed that, there is no significant structure associated with other resonances in the $\eta p$ and $p\bar{K}^0$ invariant mass distributions of Belle results. Although it may impact the fit quality to include the contributions from more resonances in the fitting, there will be more free parameters, which could enlarge the uncertainties of our results. Thus, in the present work, we do not consider the other resonances, and more complicate study could be performed when the more precise data are available.

\subsection{Invariant Mass Distributions}\label{susec:C}

There are two different amplitudes for the $N(1535)$ contribution. One is obtained by the $S$-wave pseudoscalar meson-octet baryon interactions [Eq.~(\ref{eq:amp_1535_BS})], and the other one is the Breit-Winger form [Eq.~(\ref{eq:amp_1535_BW})]. 

As we know, the $\Sigma^*(1/2^-)$ state has not yet been well confirmed experimentally and theoretically. In order to examine the existence of the $\Sigma^*(1/2^-)$ in the process $\Lambda_c^+\to pK^0_S \eta$, we first consider two models without the contribution from the $\Sigma^*(1/2^-)$. In these two cases, one is obtained with Eq.~(\ref{eq:amp_1535_BS}), denoted as `Model A,' and another one is obtained with Eq.~(\ref{eq:amp_1535_BW}, denoted as `Model B,'
\begin{eqnarray}
|\mathcal{T}^{\rm A}|^2&=&|\mathcal{T}^{N(1535)}+\mathcal{T}^{N(1650)}e^{i\phi}|^2, \label{eq:modelA}\\
|\mathcal{T}^{\rm B}|^2&=&|\tilde{\mathcal{T}}^{N(1535)}+\mathcal{T}^{N(1650)}e^{i\phi}|^2,
\end{eqnarray}
with parameter $\phi$ the relative phase between the contributions from the $N(1650)$ and the $N(1535)$.

By considering the contribution from the $\Sigma^*(1/2^-)$, we have two more models, `Model C' and  `Model D,' which are defined as follows,
\begin{eqnarray}
|\mathcal{T}^{\rm C}|^2 &=& |\mathcal{T}^{N(1535)}+\mathcal{T}^{N(1650)}e^{i\phi} \nonumber \\
& + & \mathcal{T}^{\Sigma^*(1/2^-)}e^{i\phi'}|^2,\label{eq:17}  \\
|\mathcal{T}^{\rm D}|^2 &=& |\tilde{\mathcal{T}}^{N(1535)}+\mathcal{T}^{N(1650)}e^{i\phi} \nonumber \\
&+& \mathcal{T}^{\Sigma^*(1/2^-)}e^{i\phi'}|^2, \label{eq:modelD}
\end{eqnarray} 
where $\phi'$ is a relative phase between the contributions from $\Sigma^*(1/2^-)$ and the $N(1535)$.

With all the above ingredients, one can write the double differential widths of  the process $\Lambda^+_c\to p \bar{K}^0\eta $ as follows, 
	\begin{align}
		\frac{d^2\Gamma}{dM^2_{\eta p}dM^2_{p\bar{K}^0}}=\frac{1}{(2\pi)^3}\frac{1}{32M^3_{\Lambda^+_c}}|\mathcal{T}|^2 ,\label{eq:dw1} \\
  		\frac{d^2\Gamma}{dM^2_{\eta p}dM^2_{\eta \bar{K}^0}}=\frac{1}{(2\pi)^3}\frac{1}{32M^3_{\Lambda^+_c}}|\mathcal{T}|^2 .\label{eq:dw2}
	\end{align} 
Then one could easily obtain the invariant mass squared distributions $d\Gamma/dM^2_{\eta p}$, $d\Gamma/dM^2_{p\bar{K}^0}$, and $d\Gamma/dM^2_{\eta \bar{K}^0}$ by integrating Eqs.~(\ref{eq:dw1}) and (\ref{eq:dw2}). Note that, for a given value of $M_{12}$, the range of $M^2_{23}$ is determined by~\cite{ParticleDataGroup:2022pth},
\begin{align}
	&\left(M_{23}^2\right)_{\min}=\left(E_2^*+E_3^*\right)^2-\left(\sqrt{E_2^{* 2}-m_2^2}+\sqrt{E_3^{* 2}-m_3^2}\right)^2, \nonumber\\
	&\left(M_{23}^2\right)_{\max}=\left(E_2^*+E_3^*\right)^2-\left(\sqrt{E_2^{* 2}-m_2^2}-\sqrt{E_3^{* 2}-m_3^2}\right)^2, \nonumber
\end{align}
where $E_2^{*}$ and $E_3^{*}$ are the energies of particles 2 and 3 in the $M_{12}$ rest frame, which are written as,
\begin{align}
	&E_2^{*}=\dfrac{M_{12}^2-m_1^2+m_2^2}{2M_{12}}, \nonumber\\	&E_3^{*}=\dfrac{M_{\Lambda^+_c}^2-M_{12}^2-m_3^2}{2M_{12}},
\end{align}
with $m_1$, $m_2$, and $m_3$ are the masses of involved particles 1, 2, and 3, respectively. The particle masses and widths are taken from the RPP \cite{ParticleDataGroup:2022pth}.

\section{Results and Discussions} \label{sec:Results}

\begin{table*}[htbp]
\centering
\caption{Fitted parameters of this work.}\label{tab:parameters}
\begin{tabular}{ccccccc} 
\hline\hline
Parameters         & $V_1 (\tilde{V}_1)$             & 
$V_2$              &$V_3$              &
$\phi$             &$\phi'$             &
$\chi^2/d.o.f.$         \\ 
\hline
Model A           & $13939\pm895$   &
$43369\pm3227$  & ---                   &
$(0.98\pm0.21)\pi$                  & ---                   &
5.67                  \\
Model B           & $45763\pm3421$   &
$37691\pm2817$  & ---                   &
$(0.81\pm0.16)\pi$                  &    ---                &
3.15                  \\
Model C           & $11848\pm813$   &
$52495\pm3938$  &$56230\pm4019$                    &
$(1.35\pm0.43)\pi$   &$(1.83\pm0.35)\pi$                     &
1.55                  \\
Model D           & $41614\pm2506$   &
$41026\pm2910$  &$53264\pm3248$                    &
$(1.05\pm0.31)\pi$  & $(1.60\pm0.26)\pi$                   &
1.49                  \\

\hline\hline
\end{tabular}
\end{table*}

\begin{figure*}[htb]
	\centering
	\includegraphics[scale=0.6]{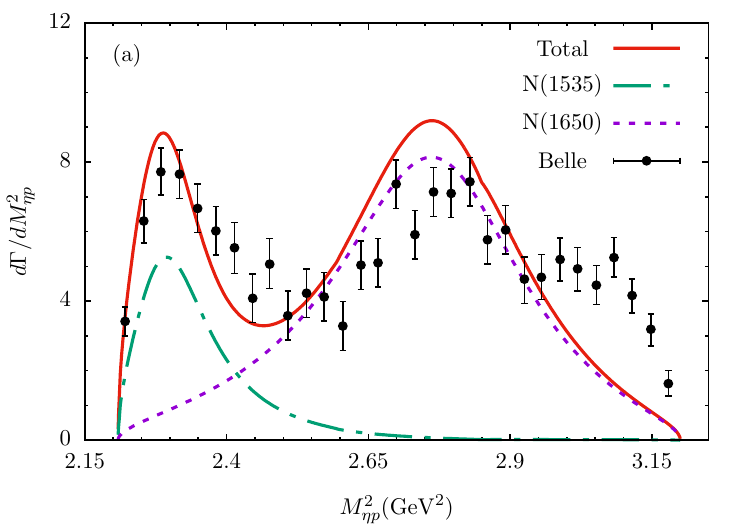}
   \includegraphics[scale=0.6]{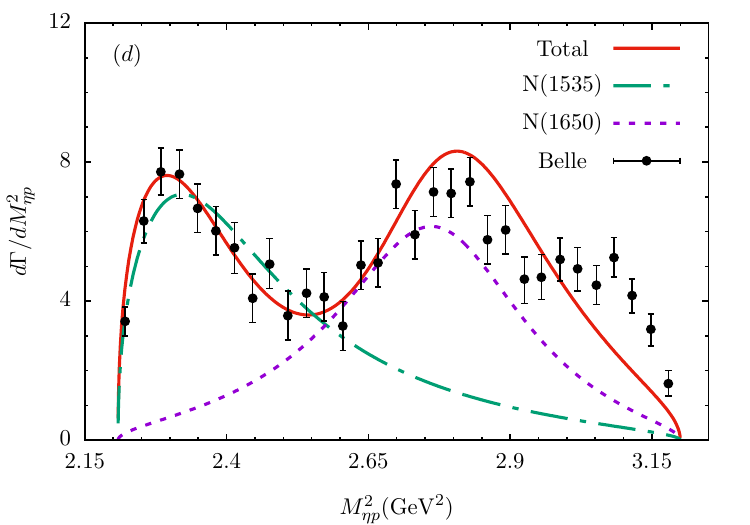}
	\includegraphics[scale=0.6]{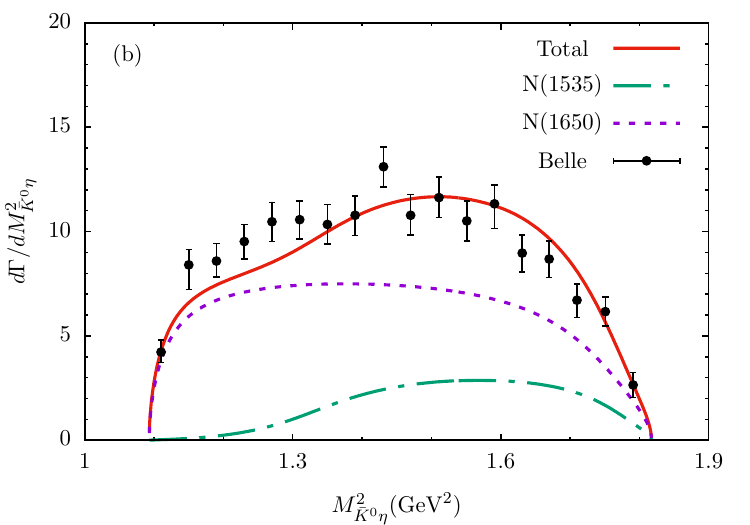}
 \includegraphics[scale=0.6]{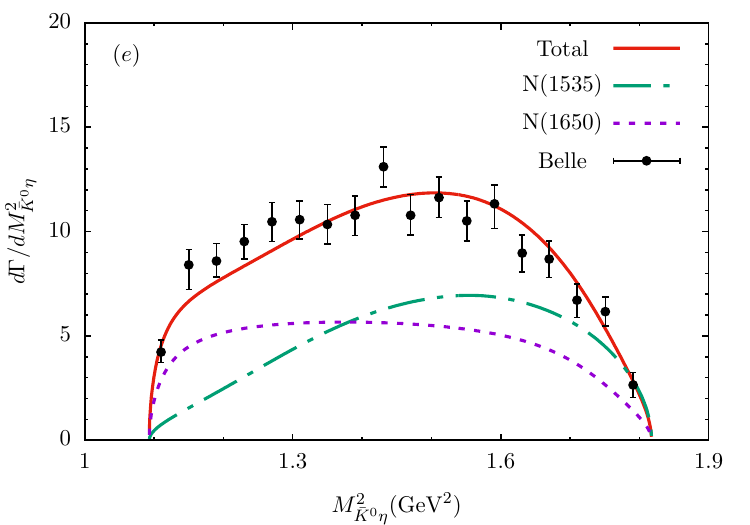}
	\includegraphics[scale=0.6]{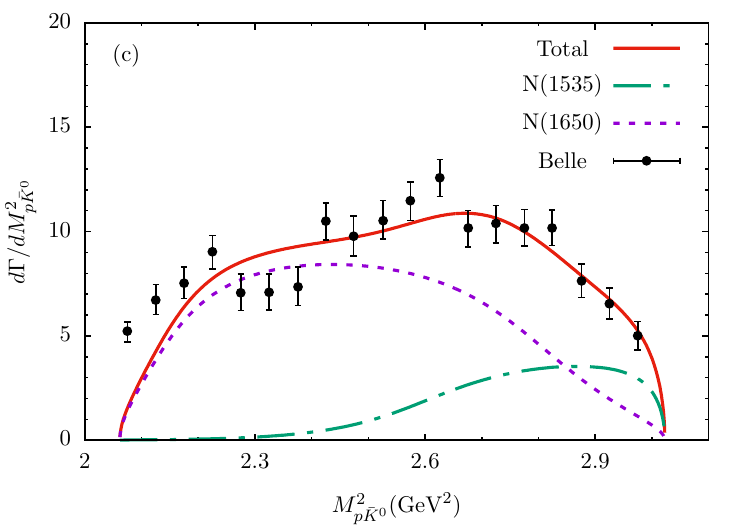}
 \includegraphics[scale=0.6]{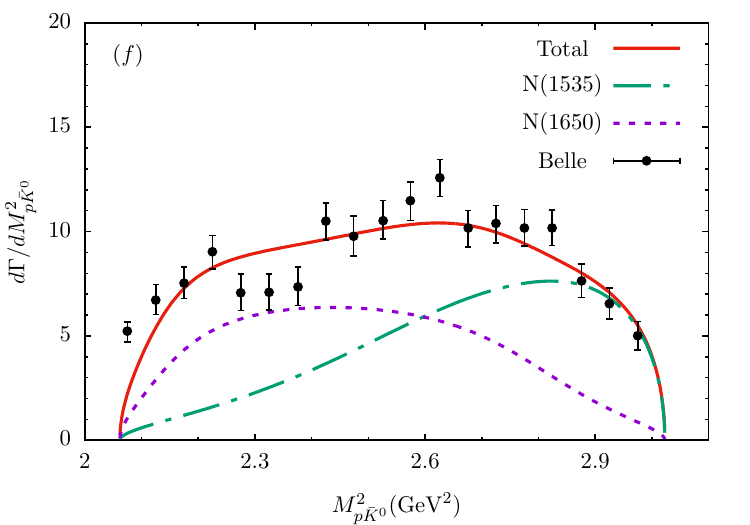}
	\caption{The $\eta p$, $\bar{K}^0\eta$, and $p\bar{K}^0$ invariant mass distributions of the process $\Lambda^+_c\to p K^0_s\eta $ with the fitted parameters of Model A (a-c) and Model B (d-f).
 The black data points with error bars labeled by `Belle' are the Belle data taken from Ref.~\cite{Belle:2022pwd},   and the solid-red curves labeled by `Total' show our theoretical results for the total contributions. In addition, we have also presented the contributions from the $N(1535)$ and $N(1650)$, which are labeled by `$N(1535)$' and `$N(1650)$', respectively. }
	\label{fig:modelAB}
\end{figure*}

 \begin{figure*}[htb]
	\centering
	\includegraphics[scale=0.6]{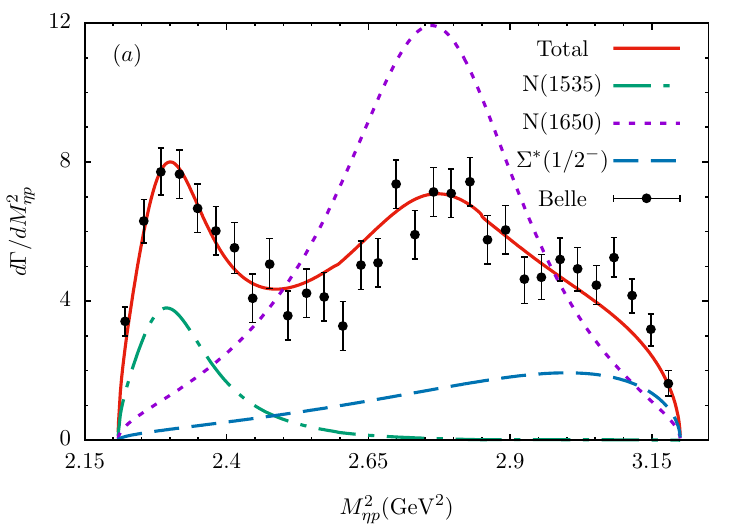}
    \includegraphics[scale=0.6]{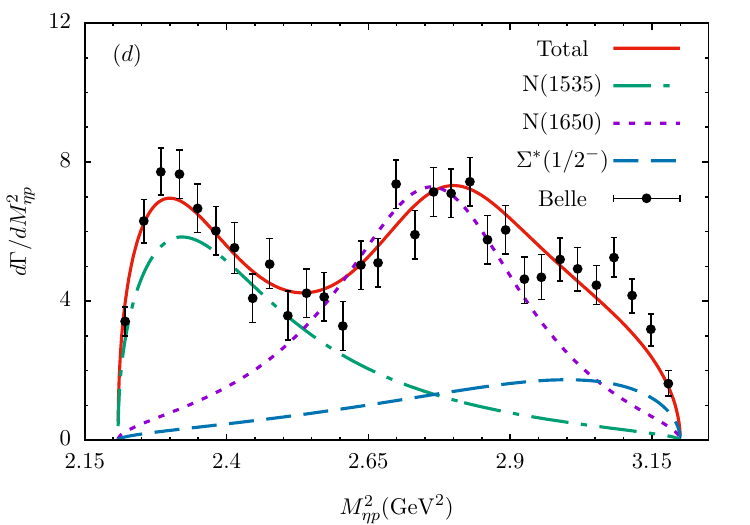}
	\includegraphics[scale=0.6]{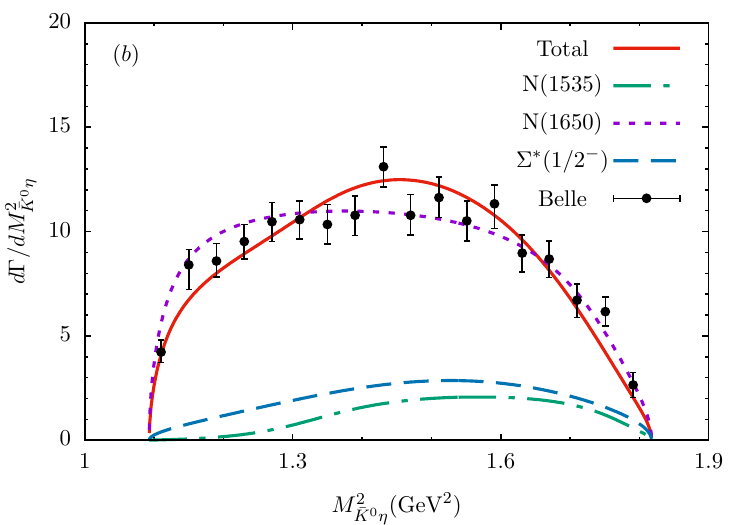}
 \includegraphics[scale=0.6]{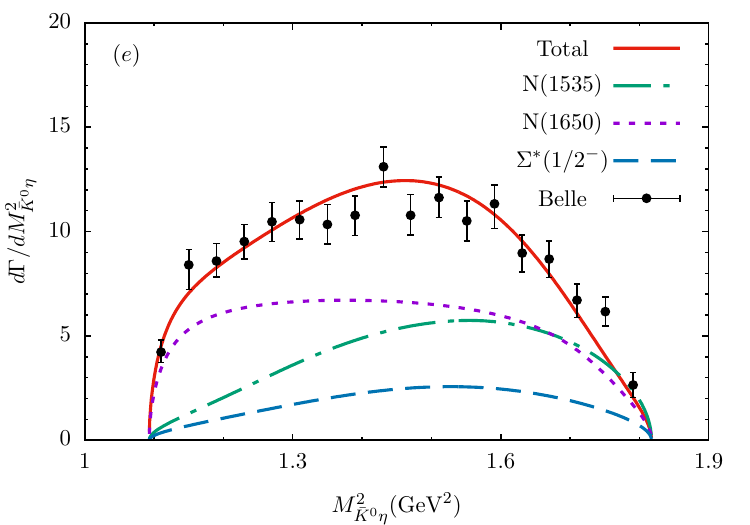}
	\includegraphics[scale=0.6]{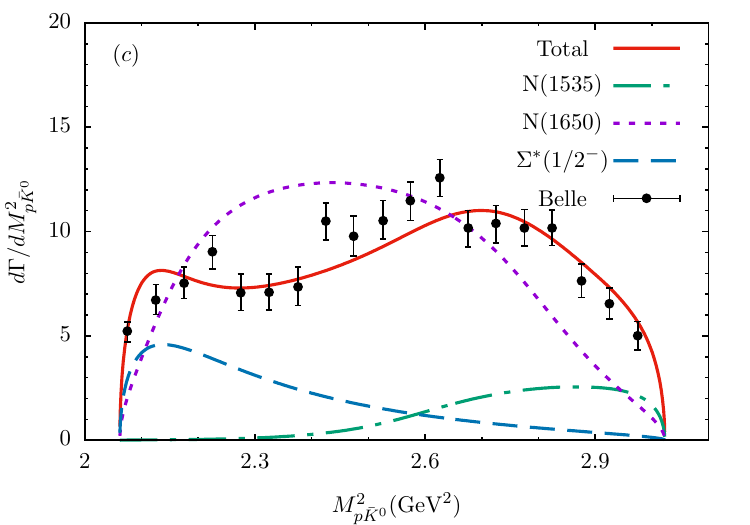}
 \includegraphics[scale=0.6]{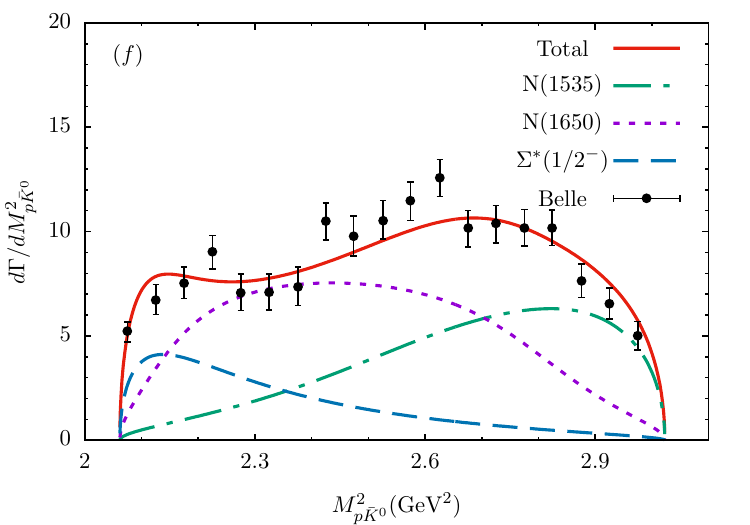}
	\caption{The $\eta p$, $\bar{K}^0\eta$, and $p\bar{K}^0$ invariant mass distributions of the process $\Lambda^+_c\to p K^0_s\eta $ with the fitted parameters of Model C (a-c) and Model D (d-f).
 The black data points with error bars labeled by `Belle' are the Belle data taken from Ref.~\cite{Belle:2022pwd}, and the solid-red curves labeled by `Total' show our theoretical results for the total contributions. In addition, we have also presented the contributions from the $N(1535)$, $N(1650)$, and $\Sigma^*(1/2^-)$, which are labeled by `$N(1535)$', `$N(1650)$', and `$\Sigma^*(1/2^-)$', respectively. }
	\label{fig:modelCD}
\end{figure*}

In our formalism, we have five free parameters, (1) $V_1 (\tilde{V}_1)$, $V_2$, $V_3$ corresponding to the strengths of the contributions from $N(1535)$, $N(1650)$, and $\Sigma^*(1/2^-)$, respectively, (2) the relative phase angle $\phi$ and $\phi'$ for the interference between different contributions from the three resonances, as shown in Eqs.~(\ref{eq:modelA})-(\ref{eq:modelD}).

First, we perform a $\chi^2$-fit of Model A and Model B to the $\eta p$, $\eta \bar{K}^0$, and $p \bar{K}^0$ invariant mass distributions of the Belle measurements~\cite{Belle:2022pwd}, which includes 67 data points in total. We have tabulated the fitted parameters and the $\chi^2/d.o.f.$ in Table~\ref{tab:parameters}. The obtained $\chi^2/d.o.f.$ is 5.67 and 3.15 for Model A and Model B, respectively. With these fitted parameters in table~\ref{tab:parameters}, we have calculated the $\eta p$, $\bar{K}^0 \eta$, and $p \bar{K}^0$ invariant mass squared distributions of the process $\Lambda_c^+\to p\bar{K}^0\eta$, as shown in Fig.~\ref{fig:modelAB}, where the left panels [Figs.~\ref{fig:modelAB}(a-c)] are the results of `Model A', and the right panels [Figs.~\ref{fig:modelAB}(d-f)] are the ones of `Model B.' One can see that although both models could reproduce the peaks of the $N(1535)$ and $N(1650)$ in the $\eta p$ invariant mass squared distribution, the high energies region of the $\eta p$ invariant mass squared distribution, the near-threshold enhancement in the $p \bar{K}^0$ invariant mass squared distribution can not be well described, which implies that other intermediate resonant contributions should be considered.

Next, we perform the fitting with Model C and Model D, involving the $\Sigma^*(1/2^-)$ contribution. The fitted parameters are tabulated in Table~\ref{tab:parameters}, and the obtained $\chi^2/d.o.f.$ are 1.55 and 1.49 for Model C and Model D, which are much smaller than the ones of Model A and Model B. With the fitted parameters, we have calculated the $\eta p$, $\bar{K}^0 \eta$, and $p \bar{K}^0$ invariant mass squared distributions of the process $\Lambda_c^+\to p\bar{K}^0\eta$, as shown in Fig.~\ref{fig:modelCD}. One finds that, including the contribution from the  $\Sigma^*(1/2^-)$, both Model C and Model D could describe the Belle measurements well. Meanwhile, according to  Figs.~\ref{fig:modelCD}(c) and \ref{fig:modelCD}(f), the near-threshold enhancement in the $p \bar{K}^0$ invariant mass squared distribution could also be well reproduced, which implies that the $\Sigma^*(1/2^-)$  plays an important role in this process.

On the other hand, in Fig.~\ref{fig:modelCD}(d), the $N(1535)$ still has a large contribution in the region above $M^2_{\eta p}>2.65$~GeV$^2$, and the theoretical results are wider than the experimental data. Furthermore, by comparing Fig.~\ref{fig:modelCD}(a) with Fig.~\ref{fig:modelCD}(d), one can easily find that  Model C could give a better description of the $N(1535)$ peak, which implies that the present experimental data favor the molecular explanation of the $N(1535)$ state.   

From the $\bar{K}^0 p$ invariant mass distributions, as shown in Figs.~\ref{fig:modelCD} (c) and (f), it is found that the contribution from the $\Sigma^*(1/2^-)$ is crucial to well describing the threshold enhancement of $\bar{K}^0 p$ invariant mass distributions measured by the Belle Collaboration. Meanwhile, by comparing the results of Model C with the ones of Model D, one can conclude that, regarding the $N(1535)$ as a dynamically generated state from the $S$-wave pseudoscalar-octet baryon interactions, the $\eta p$ invariant mass squared distribution of the Belle could be better described. 

Finally, with the fitted parameters, we could roughly estimate the ratio $R=\mathcal{B}(\Lambda_c^+\to N(1535)\bar{K}^0)/\mathcal{B}(\Lambda_c^+\to N(1650)\bar{K}^0)$ by integrating Eqs.~(\ref{eq:dw1}) and (\ref{eq:dw2}) over the invariant mass squared variables, and obtain $R=0.14\pm 0.03$ for the molecular picture of $N(1535)$, $R=0.75\pm 0.12$ for the genuine baryon state of $N(1535)$. Thus, more precise measurements about the process $\Lambda_c^+ \to p \bar{K}^0 \eta $ by BESIII, Belle II~\cite{Belle-II:2018jsg}, LHCb, and the planned Super Tau-Charm facility~\cite{Achasov:2023gey} in future could be used to test further the molecular nature of $N(1535)$, and to search for the predicted low-lying baryon $\Sigma^*(1/2^-)$.
		
\section{Summary} \label{sec:Conclusions}
    
Recently, the Belle Collaboration reported their amplitude analysis of the $\Lambda_c^+ \to p K_S^0 \eta $ and found the resonant structures corresponding to $N(1535)$ and  $N(1650)$ in the $\eta p$ invariant mass squared distribution, and a near-threshold enhancement in the $p{K}_S^0$ invariant mass squared distribution, which provides an important lab to explore the properties of the low-lying excited $N(1535)$ state and to search for the predicted $\Sigma^*(1/2^-)$ resonance.

In this work, we have investigated the process $\Lambda_c^+ \to p \bar{K}^0\eta $ by considering the contributions from the $N(1535)$, $N(1650)$, and $\Sigma^*(1/2^-)$. We have adopted two models for the contribution of the $N(1535)$; one is the molecular picture, where the $N(1535)$ is dynamically generated from the $S$-wave pseudoscalar meson-octet baryon interactions, and the other one is the Breit-Wigner form to treat the $N(1535)$ as a genuine baryon state. By fitting to the Belle measurements, one finds that the contribution from the $\Sigma^*(1/2^-)$ is crucial to well describing the Belle measurements, which implies that the present experimental data support the existence of the $\Sigma^*(1/2^-)$. 
Furthermore, although the two models with different $N(1535)$ amplitudes could give a reasonable description of the experimental $\eta p$, $\bar{K}^0 \eta$, and $p\bar{K}^0$ invariant mass squared distributions, the model regarding $N(1535)$ as s dynamically generated state could give a better description of the $N(1535)$ peak in the $\eta p$ invariant mass squared distribution. 
Furthermore, we have also estimated the ratio $R=\mathcal{B}(\Lambda_c^+\to N(1535)\bar{K}^0)/\mathcal{B}(\Lambda_c^+\to N(1650)\bar{K}^0)=0.14\pm 0.03$ and $0.75\pm 0.12$ for the molecular picture and the genuine baryon state of the $N(1535)$, respectively, which could be tested experimentally in the future, for instance in the process $\Lambda_c^+\to p\pi^0K^0_S$ with a large branching fraction of $\mathcal{B}=(1.87\pm 0.13\pm0.05) \%$~\cite{BESIII:2015bjk}.

\section*{Acknowledgments}

This work is partly supported by the National Key R\&D Program of China under Grant No. 2023YFA1606703, and No. 2024YFE0105200, and by the Natural Science Foundation of Henan under Grant No. 232300421140 and No. 222300420554, the National Natural Science Foundation of China under Grant No. 12075288, No.12475086, No. 12192263, and No. 12361141819. It is also supported by  the Open Project of Guangxi Key Laboratory of Nuclear Physics and Nuclear Technology, No. NLK 2021-08, the Central Government Guidance Funds for Local Scientific and Technological Development , China (No. Guike ZY22096024), and the Youth Innovation Promotion Association CAS.

	
\end{document}